\documentclass{article}
\usepackage[utf8]{inputenc}
\usepackage{tikz}
\usepackage{amsmath,amssymb,bbm,amsthm}
\usepackage{fullpage}
\usepackage{thm-restate,color,xcolor,xspace}
\usepackage{hyperref,cleveref}
\usepackage{algorithm}
\PassOptionsToPackage{noend}{algpseudocode}
\usepackage{algpseudocode}
\usepackage{thm-restate}
\usepackage{graphicx}
\usepackage[nice]{nicefrac}

\algrenewcommand\algorithmicrequire{\textbf{Input:}}

\usepackage{amsmath}

\usepackage{cleveref}

\newtheorem{theorem}{Theorem}[section]
\newtheorem{lemma}[theorem]{Lemma}

\newtheorem{definition}[theorem]{Definition}
\newtheorem{corollary}[theorem]{Corollary}

\newtheorem*{definition*}{Definition}
\newtheorem*{theorem*}{Theorem}

\newtheorem{claim}[theorem]{Claim}

\newtheorem{fact}[theorem]{Fact}

\def \Im {\text{Im}}
\def \Ker {\text{Ker}}
\def \wt {wt}
\def \sp {span}

\title{Unbounded Error Correcting Codes}
\author{
  Klim Efremenko\\
  \emph{Ben Gurion University}
  \and
  Or Zamir\\
  \emph{Tel Aviv University}
}
\date{}
\begin{document}

\maketitle

\begin{abstract}
 Traditional error-correcting codes (ECCs) assume a fixed message length, but many scenarios involve ongoing or indefinite transmissions where the message length is not known in advance. 
 For example, when streaming a video, the user should be able to fix a fraction of errors that occurred before any point in time.
 We introduce \emph{unbounded error-correcting codes} (unbounded codes), a natural generalization of ECCs that supports arbitrarily long messages without a predetermined length. An unbounded code with rate~$R$ and distance~$\varepsilon$ ensures that for every sufficiently large~$k$, the message prefix of length~$Rk$ can be recovered from the code prefix of length~$k$ even if an adversary corrupts up to an $\varepsilon$ fraction of the symbols in this code prefix.

We study unbounded codes over binary alphabets in the regime of small error fraction $\varepsilon$, establishing nearly tight upper and lower bounds on their optimal rate. Our main results show that:
\begin{itemize}
\item The optimal rate of unbounded codes satisfies $R<1-\Omega(\sqrt{\varepsilon})$ and $R>1-O(\sqrt{\varepsilon \log \log(1/\varepsilon)})$.

\item Surprisingly, our construction is inherently non-linear, as we prove that \emph{linear} unbounded codes achieve a strictly worse rate of $R=1-\Theta(\sqrt{\varepsilon \log(1/\varepsilon)})$.

\item In the setting of random noise, unbounded codes achieve the same optimal rate as standard ECCs, $R=1-\Theta(\varepsilon \log(1/\varepsilon))$.
\end{itemize}
These results demonstrate fundamental differences between standard and unbounded codes.
\end{abstract}

\section{Introduction}
\label{sec:ecc}

Error correction is a fundamental challenge in communication, ensuring that messages can be reliably transmitted even in the presence of noise. Standard error-correcting codes (ECCs) achieve this by encoding messages into longer structured codewords, allowing the original message to be recovered despite a bounded fraction of arbitrary errors~\cite{hamming1950error, gilbert1952comparison, varshamov1957estimate, justesen1972class, sipser1996expander}. However, standard ECCs assume a fixed message length, which does not align with many real-world scenarios involving continuous or indefinite communication.

Consider a long, continuous transmission—such as a video or audio stream—sent over an imperfect channel.
To watch, listen to, or process the stream in real time, the receiver must be able to recover from errors at any point. The ability to transmit an error-resilient stream at a high rate directly affects both buffering frequency and overall streaming performance.
In practice, such scenarios are typically handled by partitioning the communication into fixed-length \emph{packets}, each individually encoded using a standard ECC. However, this approach has a key limitation: while it ensures resilience to an $\varepsilon$-fraction of errors within each packet, it does not guarantee resilience to an $\varepsilon$-fraction of errors \emph{across the entire transmission}. This works well when errors occur randomly and independently over time but fails in the presence of bursts of noise, correlated errors, or adversarially structured errors. For example, nearly every packet may suffer slightly more than an~$\varepsilon$ fraction of errors, making it uncorrectable—even though the total fraction of errors across the entire transmission remains below~$\varepsilon$. As a result, the effective transmission rate can degrade significantly, potentially approaching~$o(1)$, as most packets remain uncorrectable.
Other practical solutions, such as packet retransmissions and feedback-based error correction protocols, rely on communication between the receiver and sender. However, beyond requiring interaction, these methods still fail to ensure a high transmission rate under all possible error patterns.
Motivated by such scenarios, we introduce a new class of codes called \emph{unbounded error-correcting codes} (unbounded codes), which extend ECCs to messages of unbounded length in a continuous manner.

An $(R,\varepsilon)$-unbounded code encodes an infinite message over an alphabet~$\Sigma$ into an infinite codeword over an alphabet~$\Gamma$ such that for any sufficiently long prefix of the codeword, the corresponding message prefix can be recovered despite an adversarial corruption of up to an $\varepsilon$ fraction of the symbols. 
More precisely, an encoding $C: \Sigma^\mathbb{N} \to \Gamma^\mathbb{N}$ is an unbounded code if for every sufficiently large $k$, the first $Rk$ symbols of the message can be uniquely determined from the first $k$ symbols of the codeword, even under worst-case noise.
For realistic applications, we do not think of the message as truly infinite, but as having an unknown and unbounded length; we then implicitly pad the message with infinitely many following zeroes. 

\begin{definition*}[Unbounded codes]
A function $C:\Sigma^{\mathbb{N}}\rightarrow \Gamma^{\mathbb{N}}$ is called an $(R,\varepsilon)$-unbounded code, if there exists~$k_0\in \mathbb{N}$ such that the following holds. 
Let $x,y\in \Sigma^{\mathbb{N}}$,~$i\geq k_0$, and ~$j\geq \frac{i}{R}$. If~$x[:i]\neq y[:i]$ then \[
d_H(C(x)[:j],C(y)[:j]) \geq \varepsilon j,
\]  
where we use~$x[:i]$ to denote the prefix of~$x$ of length~$i$, and~$d_H$ to denote the Hamming distance.
\end{definition*}

Unlike traditional ECCs, which are only defined for fixed-length messages, unbounded codes must ensure recoverability at all intermediate points. This structural constraint introduces new trade-offs between rate and error tolerance that differ from standard ECCs. We study unbounded codes over binary alphabets~$\Sigma=\Gamma=\mathbb{F}_2$ in the regime of small error fraction $\varepsilon$ and establish nearly tight bounds on their optimal rate in several natural settings.
For a comparison of optimal rates across standard and unbounded codes under various models, see Table~\ref{tab:eccs}.

\begin{table}[b]
\centering
\renewcommand{\arraystretch}{1.5}
\caption{Optimal rate~$R$ for error~$\varepsilon\rightarrow 0$.}
\label{tab:eccs}
\begin{tabular}{|l|c|c|}
\hline
                                    & \textbf{Standard ECCs}             & \textbf{Unbounded ECCs}              \\ \hline
Adversarial errors                  & $1 - \Theta(\varepsilon \log(1/\varepsilon))$ & $1 - \tilde{\Theta}(\sqrt{\varepsilon})$ \\ \hline
Random errors                       & $1 - \Theta(\varepsilon \log(1/\varepsilon))$ & $1 - \Theta(\varepsilon \log(1/\varepsilon))$ \\ \hline
Noiseless feedback                  & $1 - \Theta(\varepsilon)$               & $1 - \Theta(\varepsilon)$                 \\ \hline
\end{tabular}
\end{table}

\paragraph{Rate-Distance Tradeoff:}
We prove a nearly tight bound on the optimal rate of an unbounded code.
\begin{theorem*}
    For every small enough~$\varepsilon>0$ there exists a~$(R,\varepsilon)$-unbounded code with~$R>1-O\left(\sqrt{\varepsilon\log\log\left(1/\varepsilon\right)}\right)$. Furthermore, for every~$(R,\varepsilon)$-unbounded code it holds that~$R<1-\Omega\left(\sqrt{\varepsilon}\right)$.
\end{theorem*}
These bounds reveal that unbounded codes are less efficient than standard ECCs, which achieve $R = 1 - \Theta(\varepsilon \log(1/\varepsilon))$ in the same adversarial setting.
The construction we present to achieve this bound is non-linear, we then prove that this is inherent.

\paragraph{Separation Between Linear and Non-Linear Codes:}
Unlike standard ECCs, where linear constructions achieve optimal rates, we show that linear unbounded codes are strictly sub-optimal.
\begin{theorem*}
    For every small enough~$\varepsilon>0$ there exists a \emph{linear}~$(R,\varepsilon)$-unbounded code with rate $$R>1-O\left(\sqrt{\varepsilon\log\left(1/\varepsilon\right)}\right).$$ Furthermore, for every \emph{linear}~$(R,\varepsilon)$-unbounded code it holds that~$R<1-\Omega\left(\sqrt{\varepsilon\log\left(1/\varepsilon\right)}\right)$.
\end{theorem*}

\paragraph{Separation Between Adversarial and Random Noise Models:}
Another divergence from standard ECCs is that for unbounded codes, random and adversarial errors lead to significantly different optimal rates. In the presence of random noise (e.g., a binary symmetric channel with error probability $\varepsilon$), we show that unbounded codes achieve the same optimal rate as standard ECCs

\begin{theorem*}
    For every small enough~$\varepsilon>0$ there exists a~$(R,\varepsilon)$-unbounded code resilient to random symbol flips occurring independently with probability~$\varepsilon$ with rate~$R>1-O\left({\varepsilon\log\left(1/\varepsilon\right)}\right)$. 
\end{theorem*}

\paragraph{}
The concept of unbounded codes is partly motivated by recent questions posed by~\cite{zamir_steg} in the context of watermarking Large Language Models. In their setting, the encoder receives \emph{noiseless feedback}: each transmitted symbol is either confirmed as received correctly or flagged as corrupted in real time. Their work highlights an intriguing application of unbounded codes: imagine attempting to covertly transmit secret information by embedding it in an existing communication channel that you do not control (such as a phone call). How can you make the most of whatever transmission window you have, without knowing in advance when the connection will be cut off? In our paper, we formalize and study this problem in the more general and classical framework of coding theory, where no feedback is available and the sender receives no information about which symbols were corrupted. 

Within our proofs, we introduce and analyze a new combinatorial object we call \emph{subset codes}. While a standard error-correcting code is a set of codewords with large pairwise distance, a subset code is a collection of (potentially large) subsets of~$\mathbb{F}_2^n$, each separated by a large distance from the others — though vectors within the same subset may be arbitrarily close. These codes emerge naturally in our constructions and proofs, and in Section~\ref{sec:subset_codes} we establish tight asymptotic bounds for them. We believe subset codes may also be of independent interest.

\subsection{Connections to Prior Works}
Error-correcting codes with variable lengths have been previously studied, primarily in the context of \emph{interactive protocols}.

Our definition of unbounded codes bears some resemblance to \emph{tree codes}~\cite{schulman1993deterministic, schulman1996coding}.
As defined by Schulman, tree codes cannot have rate larger than $\frac{1}{2}$, however, recently Cohen and Samocha~\cite{CohenS20} defined a version of tree codes with rates approaching $1$.
Nonetheless, there are two crucial differences between our definition and tree codes, as well as their variants.
First, in a tree code, each codeword symbol corresponds directly to a single message symbol; in contrast, unbounded codes have no such one-to-one correspondence, and each code symbol may depend on both earlier and later message symbols. Second, the distance property in a tree code is defined only relative to the suffix of the codeword starting from the first point of disagreement between two messages. In contrast, unbounded codes define distance with respect to the entire prefix of the codeword, even if the messages initially agree and diverge only later.
For example, we require a minimum distance guarantee even when two messages differ by only their final symbol.
As a consequence, while palette-alternating tree codes achieve an optimal rate of~$1 - O(H(\varepsilon))$, unbounded codes attain the weaker rate of~$1 - \tilde{O}(\sqrt{\varepsilon})$, with each rate corresponding to a different notion of distance.

Tree codes and related constructions have been extensively studied in the context of error correction for \emph{interactive communication}~\cite{braverman2011towards,ghaffari2014optimal,gelles2017coding,efremenko2020interactive}, where the receiver can actively communicate back to the sender. This feedback allows the sender to adapt and recover from errors, making interaction a key factor in the design of coding schemes for such protocols.
In contrast, unbounded codes operate in a strictly one-way setting, where no feedback is available and the sender receives no information about which symbols were corrupted. While the optimal rates we derive for unbounded codes are similar to those achieved in interactive coding under small noise~\cite{KolR13, Haeupler14}, which are also of the form~$1-\tilde{\Theta}\left(\sqrt{\varepsilon}\right)$, we are not aware of any direct connection between these two models.

A somewhat related notion of {\em anytime capacity} \cite{SahaiM06} was studied in the context of control theory. Here, the sender does not have the whole message in advance but receives it online, and the model assumes stochastic (random) noise. The goal is that the probability of making a mistake on a bit's decoding will decrease exponentially with the time passed since the sender received this bit.

In contrast to all above, our work isolates and studies unbounded codes in a purely one-way, adversarially corrupted, no-feedback model, revealing structural and quantitative distinctions absent in these prior frameworks.

\subsection{Organization of the Paper}
In Section~\ref{sec:overview} we give a high-level overview of the constructions and proofs in the paper. In Section~\ref{sec:const} we construct linear unbounded codes and also show that their rate improves when the errors are random. In Section~\ref{sec:upperlin} we derive a simple upper bound for the rate of linear unbounded codes. In Section~\ref{sec:subset_codes} we introduce and study subset codes, which we use in the consecutive sections.
In Section~\ref{sec:upper} we present an upper bound for the rate of general unbounded codes as well as improve the bound for linear codes. 
In Section~\ref{sec:improved_const} we improve our construction using non-linear subset codes.
Finally, we conclude and present open problems in Section~\ref{sec:open}.

\section{Overview}\label{sec:overview}

\subsection{Constructions}
In Section~\ref{sec:const}, we present a simple construction of a linear unbounded code achieving~$R = 1 - O\left(\sqrt{\varepsilon \log\left(1/\varepsilon\right)}\right)$. This construction follows the spirit of classical ECC designs: each codeword symbol is formed as a random linear functional of the message symbols. The key difference is that each functional draws coefficients only for a \emph{prefix} of the message, with the remaining coefficients set to zero. We carefully choose how quickly we introduce new message symbols into these functionals to ensure that sufficient redundancy accumulates.

Intuitively, to correct a single message bit in the presence of an error rate~$\varepsilon$, we need to add roughly~$H(\varepsilon)$ bits of redundancy. Suppose we add~$\alpha$ bits of redundancy per message bit. Then, on one hand, we would need to wait for around~$H(\varepsilon)/\alpha$ additional message bits before collecting the~$H(\varepsilon)$ bits of redundancy necessary to correct that bit. On the other hand, we incur a redundancy cost of~$\alpha$ bits per message bit, limiting the achievable rate to at most~$1 - \max\left(\alpha, H(\varepsilon)/\alpha\right)$. The optimal balance is achieved by choosing~$\alpha = \sqrt{H(\varepsilon)}$.

To improve on this linear construction, we introduce the notion of \emph{subset codes} in Section~\ref{sec:subset_codes}. A subset code consists of~$K$ sets of size~$T$, with large pairwise distance between the different sets (though elements within the same set may be close). Equivalently, this structure can be viewed as an encoding~$\text{enc} : [K]\times [T] \to \mathbb{F}_2^n$ that is (a) injective and (b) satisfies that~$\text{enc}(x_1, y_1)$ and~$\text{enc}(x_2, y_2)$ are far apart for~$x_1 \neq x_2$. Thus, the encoding simultaneously encodes part of the message in a way that is robust to errors and encodes additional information that can be recovered only when no errors occur. Crucially, for large~$T$, we show that the product~$K \cdot T$ can significantly exceed the size of a standard ECC with the same distance. These constructions of set families are not linear (i.e., the sets are not linear subspaces).

In Section~\ref{sec:upper}, we use (non-linear) subset codes to construct an improved unbounded code. The construction interleaves two types of redundancy: subset-code encodings of message bits and standard checksum bits computed over the prefix of the codeword so far. This layering yields the following property: after encountering a checksum, we can correct all previously introduced errors and decode all earlier subset-code encodings fully. Between two checksums, even without full correction, the subset codes still allow partial decoding of the error-resilient portions of their inputs.

\subsection{Rate Upper Bounds}
In Section~\ref{sec:upperlin}, we establish a simpler, though not tight, upper bound on the rate of \emph{linear} unbounded codes, showing that~$R \leq 1 - \Omega(\sqrt{\varepsilon})$. The intuition is as follows. Consider the linear rank of the prefix~$C[:j]$ of the codeword. Since the rank cannot exceed the length of the prefix, for any position~$i$, the prefix~$C[:i-1]$ has rank less than~$i$. Hence, even in the absence of errors, $C[:i-1]$ cannot uniquely determine the message prefix $x[:i]$. However, by definition of unbounded codes, a longer prefix~$C[:i/R]$ suffices to recover~$x[:i]$ despite~$\varepsilon i$ errors. This implies that the segment~$C[i : i/R]$ must contain at least~$\varepsilon i$ bits devoted to redundancy (under an appropriate notion of redundancy).

Applying this reasoning repeatedly to intervals of the form~$C[Ri : i]$, $C[R^2 i : R i]$, $C[R^3 i : R^2 i]$, and so forth, we find that the first~$i$ bits of the codeword must contain approximately~$\frac{\varepsilon}{1-R} i$ bits of redundancy. The linear structure of the code ensures that this redundancy accumulates across these intervals. Yet, the code’s rate~$R$ guarantees at most~$(1 - R) i$ redundancy bits in the first~$i$ positions. Equating the two yields the inequality~$(1 - R) i \geq \frac{\varepsilon}{1 - R} i$, implying that~$1 - R \geq \sqrt{\varepsilon}$.

In Section~\ref{sec:improved_const}, we extend this rate upper bound to general (non-linear) unbounded codes and also refine the bound in the linear setting. We follow a similar proof strategy, but in place of linear algebraic arguments, we use information-theoretic tools. Instead of analyzing ranks of codeword prefixes, we reason about their entropy. In the non-linear case, codeword bits may simultaneously carry error-resilient information about some message bits and non-resilient information about others that cannot yet be decoded. This behavior is captured naturally by the structure of subset codes. Accordingly, we also apply bounds on subset codes, derived in Section~\ref{sec:subset_codes} using isoperimetric inequalities on the hypercube. Stronger lower bounds for linear subset codes allow us to obtain a tight rate upper bound of~$(1 - R) \geq \Omega\left(\sqrt{\varepsilon \log\left(1/\varepsilon\right)}\right)$ for linear unbounded codes over~$\mathbb{F}_2$, and recover the rate upper bound of~$(1 - R) \geq \Omega(\sqrt{\varepsilon})$ in the setting of general codes.

\section{Preliminaries}\label{sec:prelim}
We use the (Shannon) entropy function throughout the paper; for a discrete random variable~$X$ distributed according to~$p(x)$, it is defined as~$H(X):=\sum_{x\in \text{Support}(X)} -p(x) \log p(x)$. For a detailed reminder of basic properties of entropy and information theory see for example the book of Gray~\cite{gray2011entropy}. We often abuse notation and denote by~$H(x):=-x \log x - (1-x) \log (1-x)$ the binary entropy function. Note that 
\[
\frac{d}{dx} H(x) = \log (1-x) - \log x, \ \ \ \frac{d^2}{dx^2} H(x) = -\frac{1}{x(1-x)\ln 2}
.
\]
We also make frequent use of the following helpful corollary of Stirling's approximation:
\begin{equation*}\label{eq:stirling}
    \binom{n}{\alpha n} = \Theta^*(2^{H(\alpha)n}); \text{ or, equivalently,  }
    \binom{n}{\alpha n} = 2^{H(\alpha)n} \cdot 2^{\Theta(\log n)}.
\end{equation*}

\section{Construction of a Linear Unbounded Code}\label{sec:const}

We give a non-explicit construction of a linear code, similar to the classic Gilbert-Varshamov construction~\cite{gilbert1952comparison,varshamov1957estimate}.
The two significant differences between the classic construction and the one for unbounded codes are as follows.
First, as the number of message coefficients is unbounded, a code coefficient can clearly only be a function of a finite subset of them.
Second, as the previous point implies that message coefficients will begin affecting the code coefficients gradually, this also means that when a message coefficient begins affecting the code coefficients it will be initially involved in too few code coefficients to be recoverable after the addition of errors.
We solve those problems by choosing the code coefficients to be random linear functions over gradually-increasing prefixes of the message coefficients, and by analysing the distance in a more delicate manner.

\begin{theorem}\label{thm:const}
    For every~$\varepsilon < \frac{1}{17}$ and~$R<1-4\sqrt{\varepsilon\log \frac{1}{\varepsilon}}$, there exists a linear $(R,\varepsilon)$-unbounded code.
\end{theorem}

Let~$\varepsilon>0$ be a small enough positive number, and fix~$R\in(0,1)$ to be chosen later. Another parameter to be chosen later is~$\tau \in (R,1)$, which will be the rate of variable introduction.
We define the code~$C$ as a random variable as follows.
The~$i$-th code word coefficient~$C_i := C(x)_i$ is drawn as a uniformly chosen linear functional over the first~$\lceil \tau i \rceil$ message coefficients~$x[1],x[2],\ldots,x[\lceil \tau i \rceil]$.

\begin{lemma}\label{lem:const_tau}
    If~$\tau + H\left(\frac{\varepsilon}{1-{R}/{\tau}}\right) < 1$, then~$C$ is a~$(R,\varepsilon)$-unbounded code with positive probability.
\end{lemma}

Denote by~$B_{i,j}$ for some~$i$ and~$j>i/R$, the event that there exist two message words~$y_1,y_2$ such that~$i$ is the smallest index for which~$y_1[i]\neq y_2[i]$, and that~$d_H(C(y_1)[:j],C(y_2)[:j]) < \varepsilon j$.

\begin{claim}\label{claim:const_bij}
    For every~$i,j$ for which~$B_{i,j}$ is defined,
    $$
    \log \Pr\left(B_{i,j}\right) \leq \left(\tau + H\left(\frac{\varepsilon}{1-{R}/{\tau}}\right)-1\right)\left(1-\frac{R}{\tau}\right)j+o(j)
    .$$
\end{claim}
\begin{proof}
Denote by~$k:=\lceil \tau j\rceil$.
By definition, the code word coefficients~$C_1,\ldots, C_j$ are linear functionals supported only on the message coefficients~$x[1],\ldots, x[k]$.
If~$y_1,y_2$ as in the definition of~$B_{i,j}$ exist, then denote by~$x=y_2-y_1$, this is a vector for which the first non-zero coordinate is~$x[i]$, and for which~$\wt \left(C\left(x\right)[:j]\right) < \varepsilon j$.
Let~$x$ be a vector of length~$k$ in which the first non-zero coordinate is~$x[i]$.
For every~$\frac{i}{\tau} \leq \ell \leq j$, the functional~$C_{\ell}$ includes~$x[i]$ with probability~$\frac{1}{2}$, and hence~$C_{\ell}(x)\neq 0$ with probability~$\frac{1}{2}$, and those are independent for different indices~$\ell$. The number of such functionals~$C_{\ell}$ is~$j':=j-\lceil\frac{i}{\tau}\rceil+1 \geq j-\frac{i}{\tau}\geq j-\frac{Rj}{\tau} = \left(1-\frac{R}{\tau}\right)j$.
In particular, taking a union bound we get that
\[
\Pr\left(\wt\left(C\left(x\right)\left[:j\right]\right) < \varepsilon j\right) \leq
{j' \choose <\varepsilon j} 2^{-j'} 
\leq
2^{\left(H\left(\varepsilon j/{j'}\right)-1\right)j'+o(j)}
.
\]
Such a vector~$x$ has only~$k-i$ coefficients that are not predetermined by the definition, and hence another union bound shows that
\[
\Pr\left(B_{i,j}\right) \leq
2^{k-i} \cdot 2^{\left(H\left(\varepsilon j/{j'}\right)-1\right)j'+o(j)}.
\]
We next observe that~$k-i = \lceil \tau j \rceil - i \leq \tau j - i + 1 \leq \tau j' + 1$, and also that~$\varepsilon j / j' \leq \varepsilon / \left(1-\frac{R}{\tau}\right)$. We conclude that
\begin{align*}
\log \Pr\left(B_{i,j}\right) &\leq
\tau j' + \left(H\left(\frac{\varepsilon}{1-{R}/{\tau}}\right)-1\right)j'+o(j)\\
&\leq
\left(\tau + H\left(\frac{\varepsilon}{1-{R}/{\tau}}\right)-1\right)j'+o(j).
\end{align*}
In the settings of Lemma~\ref{lem:const_tau}, the expression in the outermost parenthesis is negative, and thus we may use the previous inequality~$j'\geq \left(1-\frac{R}{\tau}\right)j$.
\end{proof}

\begin{proof}[Proof of Lemma~\ref{lem:const_tau}]
Denote by~$\alpha := -\left(\tau + H\left(\frac{\varepsilon}{1-{R}/{\tau}}\right)-1\right)\left(1-\frac{R}{\tau}\right)$, in our settings~$\alpha>0$.
By Claim~\ref{claim:const_bij}, $\Pr\left(B_{i,j}\right)\leq 2^{-\alpha j + o(j)}$ whenever the event is defined.
By a union bound, the probability that~$C$ is \emph{not} a~$(R,\varepsilon)$-unbounded code with respect to a minimum length~$k_0$, is at most
$$
\sum_{i=k_0}^{\infty} \sum_{j > \frac{i}{R}}^{\infty}  2^{-\alpha j + o(j)}
\leq
\sum_{i=k_0}^{\infty} \frac{1}{1-2^{-\alpha}}  2^{-\alpha {i}/{R} + o(i/R)}
\leq
\frac{1}{\left(1-2^{-\alpha}\right)\left(1-2^{-\alpha/R}\right)}
2^{-\alpha {k_0}/{R} + o(k_0/R)}
.$$
In particular, this probability is strictly smaller than~$1$ for a large enough constant~$k_0$ (that depends only on~$R,\varepsilon$).
\end{proof}

\begin{proof}[Proof of Theorem~\ref{thm:const}]
    If there exists~$\tau$ for which the condition of Lemma~\ref{lem:const_tau} is satisfied, then a~$(R,\varepsilon)$-unbounded code exists as well.
    Let~$\varepsilon<$ be a small enough constant, and let~$R=1-4\sqrt{\varepsilon\log \frac{1}{\varepsilon}}$. We show that~$\tau=\frac{R}{1-\sqrt{\varepsilon\log {1}/{\varepsilon}}}$ satisfies the condition.
    We first observe that 
    $$\tau=\frac{R}{1-\sqrt{\varepsilon\log \frac{1}{\varepsilon}}}  
    =\frac{1-4\sqrt{\varepsilon\log \frac{1}{\varepsilon}}}{1-\sqrt{\varepsilon\log \frac{1}{\varepsilon}}} <
    \left(1-4\sqrt{\varepsilon\log \frac{1}{\varepsilon}}\right)  \left(1+2\sqrt{\varepsilon\log \frac{1}{\varepsilon}}\right)
    <
    1-2\sqrt{\varepsilon\log \frac{1}{\varepsilon}}
    ,$$
    where the first inequality follows as~$\frac{1}{1-z} < 1+2z$ for all~$0<z<\frac{1}{2}$. 
    We then see that
    $$
    H\left(\frac{\varepsilon}{1-{R}/{\tau}}\right) =
    H\left(\sqrt{\frac{\varepsilon}{\log \frac{1}{\varepsilon}}}\right) <
    2 \sqrt{\frac{\varepsilon}{\log \frac{1}{\varepsilon}}} \log \left(\sqrt{\frac{\log \frac{1}{\varepsilon}}{\varepsilon}}\right) <
    2\sqrt{\varepsilon\log \frac{1}{\varepsilon}}
    ,$$
    where the first inequality follows as~$H(z)<2z\log(1/z)$ for all~$0<z<\frac{1}{2}$, and the second inequality follows as~$\sqrt{\frac{\log \frac{1}{\varepsilon}}{\varepsilon}} < \frac{1}{\varepsilon}$ for every~$\varepsilon>0$.
    We therefore satisfy the conditions of Lemma~\ref{lem:const_tau}.
\end{proof}

\subsection{Random Errors}
The construction and analysis in the presence of random (rather than adversarial) errors is similar. We make use of the fact that now the errors are spread uniformly and in particular a suffix of the code is expected to only contain~$\varepsilon$ fraction of errors relative to its size, rather than possibly containing~$\varepsilon n$ errors where~$n$ is the size of the entire code-word.

Let us start with the definition of unbounded codes w.r.t to random errors. 
\begin{definition}[Unbounded codes for random errors]
An encoding $C:\mathbb{F}_2^{\mathbb{N}}\rightarrow \mathbb{F}_2^{\mathbb{N}}$ is called an \emph{unbounded code} with rate~$R$ resilient to $BSC(\varepsilon)$ noise
 if there exist~$k_0\in \mathbb{N}, c>0$ and a decoder
$D:\mathbb{F}_2^{\mathbb{N}}\rightarrow \mathbb{F}_2^{\mathbb{N}}$ such that the following holds. 
Let $x\in \mathbb{F}_2^{\mathbb{N}}$,~$i\geq k_0$, and~$j\geq \frac{i}{R}$, then
$$\Pr\left( D(C(x)[:j]\oplus \text{Ber}(\varepsilon)^j)[:i]\neq  x[:i]\right)\leq \exp(-c j)$$. 
\end{definition}

\begin{theorem}\label{thm:const_rand}
    For every~$\varepsilon < \frac{1}{17}$ and~$R<1-H(3\varepsilon)$, there exists a linear $(R,\varepsilon)$-unbounded code resilient to $BSC(\varepsilon)$ noise. 
 \end{theorem}

To avoid repetition, we defer the proof to Appendix~\ref{appendix:rand_errors}.

\section{Simple Upper Bound for Linear Codes}\label{sec:upperlin}
A unbounded code~$C$ is called \emph{linear} if the code function~$C$ is a linear function.
\begin{theorem}\label{thm:upperlin}
    For every linear $(R,\varepsilon)$-unbounded code it holds that $R\leq 1-\sqrt{\varepsilon}$.
\end{theorem}

Let~$C:\Sigma^{\mathbb{N}}\rightarrow \Sigma^{\mathbb{N}}$ be a linear code.
Denote by~$P_i : \Sigma^{\mathbb{N}} \rightarrow \Sigma^i$ be the projection to the first~$i$ coordinates.
Denote by~$H(i) := \dim \left(\Im \left(P_i \circ C\right)\right)$ the dimension of the projection of~$C$ to the first~$i$ coordinates.
We observe that~$H(0)=0$ and~$H(i+1)\leq H(i)+1$ for every~$i$, hence also~$H(i)\leq i$.
For every~$i$, let~$H^{-1}(t)$ be the minimal integer~$i$ such that~$H(i)\geq t$.
For every~$i$ we also denote by~$r(i):=i-H(i)\geq 0$ the \emph{redundancy} of the~$i$-prefix of the code~$C$.

\begin{lemma}\label{lem:r_upper}
    Let~$C$ be a~$(R,\varepsilon)$-unbounded linear code.
    For every~$i\geq k_0$ we have~$r(i)\leq (1-R)i + 1$.
\end{lemma}
\begin{proof}
    Assume that~$r(i)>(1-R)i + 1$.
    Hence,~$H(i) = i - r(i) < Ri - 1$.
    On the other hand, by the definition of a distance~$R$ (unbounded) code, all possible messages of length~$\lfloor Ri \rfloor$ must have distinct code prefixes of length~$i$, and thus~$H(i)\geq \lfloor Ri\rfloor \geq Ri - 1$, which is a contradiction.
\end{proof}

\begin{lemma}\label{lem:linearsub}
    Let~$C$ be a~$(R,\varepsilon)$-unbounded linear code.
    For every~$i,j\in \mathbb{N}$ with~$i\geq k_0$ and~$j\geq\frac{i}{R}$ it holds that
    \[
    r\left(j\right) - r\left(H^{-1}\left(i\right)\right) \geq \varepsilon j - 1
    .
    \]
\end{lemma}
\begin{proof}
    Since~$C$ is linear and is a~$(R,\varepsilon)$-unbounded code, for every~$x$ such that~$x[1:i]\neq 0$ we have~$$\wt\left(C\left(x\right)\left[1:j\right]\right) \geq \varepsilon j.$$
    Denote by~$L=\sp \left( \{e_1,\ldots,e_i\} \right) \cong \Im P_i$ be the linear subspace of vectors whose support is in the first~$i$ coordinates. We have~$\dim L = i$.
    For~$k$ to be picked later, consider the linear map~$P_k \circ C$ applied to~$L$.
    As a restriction to a subspace, we have~$\dim \Im_L \left(P_k \circ C\right) \leq \dim \Im \left(P_k \circ C\right) = H(k)$.
    Hence,~$\dim \Ker_L \left(P_k \circ C\right) \geq i - H(k)$.
    In particular, we set~$k:=H^{-1}(i)-1$ and thus~$H(k)=i-1$, and~$\dim \Ker_L \left(P_k \circ C\right) \geq 1$. We conclude that there exists~$0\neq v_0 \in L$ such that~$P_k\left(C\left(v_0\right)\right)=0$. This is a non zero message vector~$v_0$ of length at most~$i$, with a corresponding code word that begins with~$k$ zeros.

    Next, consider the kernel of~$P_k$ when applied to~$P_{j} \circ C$, this is the linear subspace~$T$ of all code-word prefixes of length~$j$ in which the first~$k$ coordinates are zero. 
    As~$P_k\circ P_j \circ C = P_k \circ C$ and as~$\dim \Im \left(P_k \circ C\right) = H(k)$, we have that~$\dim T = H(j)-H(k) = H(j) - i + 1$.
    There is a subset~$S\subseteq [k+1:j]$ of~$|S|=\dim T$ coordinates such that~$T$ projected to~$S$ is of full rank.
    In particular, there exists a linear basis~$t_1,\ldots,t_{\dim T}$ of~$T$ such that every~$t_j$ has zeros in all but one coordinate of~$S$.
    If there exists any~$j$ for which~$C^{-1}(t_j)[1:i]\neq 0$ then denote by~$v=C^{-1}(t_j)$.
    Otherwise, consider the unique linear combination~$y:= C(v_0)+\sum_j \alpha_j t_j$ in which all the coordinates of~$S$ are zero. As we have~$v_0[1:i]\neq 0$ and~$C^{-1}(t_j)[1:i] = 0$ for every~$j$, then we may set~$v:=C^{-1}(y)$ and have~$v[1:i]\neq 0$.
    
    We constructed a message vector~$v$ such that~$v[1:i]\neq 0$, $C(v)[1:k]=0$ and~$\wt\left(C\left(v\right)\vert_{S}\right)\leq 1$.
    We conclude that
    \begin{align*}
    \varepsilon j \leq \wt\left(C\left(v\right)[1:j]\right) &\leq
    j-k-\dim T + 1 \\
    &= j-\left(H^{-1}(i)-1\right) - \left(H\left(j\right)-i+1\right) + 1\\
    &= \left(j-H\left(j\right)\right) - \left(H^{-1}\left(i\right) - i\right) + 1 \\
    &=r\left(j\right) - r\left(H^{-1}\left(i\right)\right) + 1.
    \end{align*}
\end{proof}

\begin{proof}[Proof of Theorem~\ref{thm:upperlin}]
Let~$n',n\in \mathbb{N}$ be large enough integers, we think of~$n$ as much larger than~$n'$, and of~$n'$ as a large constant. When we write~$o(1)$ or~$\omega(1)$ throughout the proof, it is with respect to~$n'\rightarrow\infty$ and~$\frac{n}{n'}\rightarrow \infty$. We define the sequence~$n_0,\ldots n_k$ recursively by~$n_0:=n$, then $n_{k+1} := H^{-1}\left(\lfloor Rn_k \rfloor\right)$ for every~$k\geq 0$, and finally~$K$ is the largest~$k$ such that~$n_k > n'$.
For every~$0\leq k < K$ by using Lemma~\ref{lem:linearsub} with~$j=n_k,i=\lfloor Rn_k \rfloor$, we have~$r(n_k)-r(n_{k+1}) \geq \varepsilon n_k - 1$.
Also for every~$k$, we have~$n_{k+1} = H^{-1}\left(\lfloor Rn_k \rfloor\right) \geq \lfloor Rn_k \rfloor \geq Rn_k - 1$. 
By iterative application of the previous inequality, we also have~$n_k \geq R^k n_0 - \frac{1}{1-R}$. This also implies that~$K= \omega(1)$.
Consider the following telescopic summation,
\begin{align*}
r\left(n_0\right) - r\left(n_{K}\right) 
=
\sum_{k=0}^{K-1} \left(r\left(n_k\right) - r\left(n_{k+1}\right)\right)
&\geq
\sum_{k=0}^{K-1} \left(\varepsilon n_k - 1\right)\\
&=
\left(1-o\left(1\right)\right) \varepsilon \sum_{k=0}^{K-1} n_k\\
&\geq
\left(1-o\left(1\right)\right) \varepsilon \sum_{k=0}^{K-1} \left(R^k n_0 - \frac{1}{1-R}\right)\\
&=
\left(1-o\left(1\right)\right)  \varepsilon n \sum_{k=0}^{K-1} R^k \\
&=
\left(1-o\left(1\right)\right)  \frac{(1-R^K)\varepsilon n}{1-R}\\
&=
\left(1-o\left(1\right)\right)  \frac{\varepsilon n}{1-R}
.
\end{align*}
We finally observe that~$r(n_K)\geq 0$, and that by Lemma~\ref{lem:r_upper} we also have~$r(n_0)\leq (1-R)n+1$.
We conclude with the following chain of inequalities, beginning with the above.
\begin{align*}
    r(n_0)-r(n_K) &\geq \left(1-o\left(1\right)\right)  \frac{\varepsilon n}{1-R} \\
    (1-R)n+1 &\geq \left(1-o\left(1\right)\right)  \frac{\varepsilon n}{1-R}
    \\
    (1-R)^2 &\geq \left(1-o\left(1\right)\right)  \varepsilon
    \\
    1-R &\geq \left(1-o\left(1\right)\right)  \sqrt{\varepsilon},
\end{align*}
And as~$R,\varepsilon$ are fixed constants, this implies that~$R\leq 1-\sqrt{\varepsilon}$ as desired.
\end{proof}

\section{Subset Codes}\label{sec:subset_codes}
In both the upper bound and construction for non-linear codes we make use of a combinatorial object we call \emph{Subset Codes}. An error correcting code is a set of points in~$\mathbb{F}_2^n$ that are far apart from each other, a \emph{subset code} is a set of \emph{subsets} of~$\mathbb{F}_2^n$ such that each two such subsets are far from each other.

\begin{definition}
    A $(K,T,\delta,n)$-subset code is a collection~$\mathcal{S}=\{S_1,\ldots,S_K\}$ of subsets~$S_i\subset \mathbb{F}_2^n$ such that every subset is of size~$|S_i|\geq T$ and moreover for every~$i\neq j$,~$x\in S_i$ and~$y\in S_j$ we have~$|x-y|>\delta n$.
\end{definition}

Or equivalently, one can define it as 
\begin{definition}
    A $(K,T,\delta,n)$-subset code is a mapping $C:[K]\times [T] \rightarrow \mathbb{F}_2^n$ such that $C$ is one-to-one and for $x_1 \neq x_2$ and any $y_1, y_2$ it holds that $|C(x_1,y_1)-C(x_2,y_2)|\geq \delta n$. 
\end{definition}

For example, we can construct a~$\left(2,{n \choose <\left(\frac{1-\delta}{2}\right)n},\delta,n\right)$-subset code by taking one subset to be all vectors with weight~$< \frac{1-\delta}{2}n$ and the other to be all vectors with weight~$> \frac{1+\delta}{2}n$.
We observe that a subset code gives us an interesting type of encoding: Consider the encoding function~$e:[K]\times [T] \rightarrow \mathbb{F}_2^{n}$, that maps a pair~$k\in[K],t\in[T]$ into the~$t$-th element of~$S_k$, according to some consistent orderings. 
Due to the distance property of subset codes, even if we add~$\frac{\delta}{2} n$ errors to~$e(k,t)$, we can still recover~$k$; On the other hand, if there are no errors at all, we can recover both~$k$ and~$t$ from~$e(k,t)$ as it is injective.
Hence, a subset-code gives us an encoding that allows us to recover a certain number of message bits if there were errors added to the codeword, and a larger number of message bits if there were no errors added to the codeword. 
The simple construction sketched in the beginning of this paragraph, for example, allows us to recover~$1$ bit in the presence of~$\delta/2$ errors, and~$(1-\theta(\delta^2))n$ bits if there are no errors present.

Next, we give upper and lower bounds for the best parameters possible in a subset code. 
The following isoperimetric inequality for vertices in the hypercube is due to Harper (see~\cite{harper1966optimal,bollobas1986combinatorics,calabro2004harper,raty2020uniqueness}).

\begin{theorem}[Harper's Inequality]
    For every~$m,r_1,r_2\in \mathbb{N}$ such that~$r_1+r_2 \leq m$, and any~$A\subseteq \mathbb{F}_2^m$ of size~$|A|\geq |B(r_1)|$, we have~$|A+B(r_2)|\geq |B(r_1+r_2)|$.
    Here~$B(r)$ is the Hamming ball of radius~$r$ in~$\mathbb{F}_2^n$.
\end{theorem}

As a corollary of Harper's inequality, we deduce the following.

\begin{lemma}\label{lem:iso_many}
    Let~$k\leq n$, and let~$r$ be the minimal radius such that the size of the radius~$r$ ball in~$\mathbb{F}_2^n$ is at least~$|B(r)| \geq 2^{n-k}$.
    Then, for any~$\delta>0$ there is no~$\left(2^k, {n \choose \leq \lceil r - \delta n/2\rceil+2}, \delta, n\right)$-subset code.
\end{lemma}
\begin{proof}
    Let~$B_{\delta/2}:=B\left(\lfloor\delta n/2\rfloor\right)$ be the Hamming ball of radius~$\lfloor\delta n/2\rfloor$ around~$0$ in~$\mathbb{F}_2^n$.
    By Harper's Inequality, for any code subset~$S_i\in \mathcal{S}$, $$|S_i+B_{\delta/2}| \geq |B\left(\lceil r - \delta n/2\rceil+2+\lfloor\delta n/2\rfloor\right)|\geq |B\left(r+1\right)| > 2^{n-k}.$$
    As~$\sum_{i=1}^{2^k}|S_i+B_{\delta/2}| > |\mathbb{F}_2^n|$, there must be~$i_1\neq i_2$ such that~$S_{i_1}+B_{\delta/2}$ intersects~$S_{i_2}+B_{\delta/2}$. In particular there are~$x\in S_{i_1}$ and~$y\in S_{i_2}$ of distance at most~$2\lfloor\delta n/2\rfloor \leq \delta n$ from each other.
\end{proof}

Denote by~$\alpha=1-\frac{k}{n}$, note that~$\frac{r}{n} = H^{-1}\left(\alpha\right)+o(1)$ as~$n\rightarrow \infty$.
We also note that due to the concavity of the binary entropy function,~$H(x-\delta)\leq H(x)-H'(x)\cdot\delta$. We furthermore note that~$H'(x)=\Theta\left(\log\left(1/x\right)\right)$. We hence have
\[
{n \choose \leq \lceil r - \delta n/2\rceil+2} = 
2^{H\left(\frac{r}{n} - \frac{\delta}{2}\right)n + o(n)} \leq
2^{H\left(\frac{r}{n}\right)n - H'\left(\frac{r}{n}\right)\frac{\delta}{2}n + o(n)} =
2^{\alpha n} \cdot 2^{-\Theta\left(\delta \log\left(n/r\right)\right)n + o(n)}.
\]
We also note that~$\log\left(\frac{n}{r}\right) = \Theta\left(\log\frac{1}{\alpha}\right)$.
We thus conclude from~\ref{lem:iso_many} the following.
\begin{corollary}\label{cor:no_subsetcode}
    For any~$\alpha,\delta>0$, there is no
    ~$\left(2^{\left(1-\alpha\right) n},2^{\left(\alpha-\Theta\left(\delta \log\left(1/\alpha\right)\right)\right)n},\delta,n\right)$-subset code for all large enough~$n$.
\end{corollary}

We next show that the above bound is asymptotically tight by constructing a subset code with similar parameters.
\begin{lemma}
    Let~$k\leq n$, and let~$r$ be the maximal radius such that the size of the radius~$r$ ball in~$\mathbb{F}_2^n$ is at most~$|B(r)| \leq 2^{n-k}$.
    Then, for any~$\delta>0$ there exists a~$\left(2^{k-1}, \frac{1}{2}{n \choose \leq \lfloor r - \delta n\rfloor}, \delta, n\right)$-subset code.
   \end{lemma}
\begin{proof}
    We construct the subset code greedily with the following algorithm.
    \begin{enumerate}
        \item Set $\mathcal{S}_0=\emptyset$, $\mathcal{P}_0=\mathbb{F}_2^n$.
        \item For $i=1, \ldots,  2^{k-1}$ do:
        \begin{enumerate}
            \item{\label{alg:pick}} Pick point $s_i$ such that $B(s_i,r-\delta n) \cap \mathcal{P}_{i-1} \geq |B(r-\delta n)|/2$.
            \item  Denote by $S_i=B(s_i,r-\delta n) \cap \mathcal{P}_{i-1}$. 
            \item  Set $\mathcal{P}_{i}=\mathcal{P}_{i-1}\setminus B(s_i,r)$ and~$\mathcal{S}_i = \mathcal{S}_{i-1} \cup \{S_i\}$.
        \end{enumerate}
    \end{enumerate}

Note that~$|B(r)|\leq 2^{n-k}$ and hence~$|\mathcal{P}_i|\geq 2^n - 2^{n-k}\cdot i \geq 2^{n-1}$.
In particular, for a random point $s_i$ the expected size of $B(s_i, r-\delta n) \cap \mathcal{P}_{i-1}$ is at least~$|B(r-\delta n)|/2$ and therefore we can always execute line number \ref{alg:pick} in the greedy algorithm.

Also, note that for every $x\in S_i$ and $y\in \mathcal{P}_{i}$ it holds that $d_H(s_i, x) \leq r-\delta n$ and $d_H(s_i, y) \geq r$ therefore by triangle inequality $d_H(x, y) \geq \delta n$. Since for all $j>i, S_j \subset \mathcal{P}_{i}$ it holds that the distance of the subset code~$\mathcal{S}=\mathcal{S}_{2^{k-1}}$ is at least~$\delta n$.
\end{proof}

As before, we analyze the asymptotic of the parameters in the construction.
Note that for any~$x \in(0,1/2)$ we have~$H(x-\delta)\geq H(x) - H'(x)\delta - \Theta\left(\frac{\delta^2}{x}\right)$. We conclude the following.

\begin{corollary}\label{cor:subset}
    For any~$\alpha>0$ and $0<\delta\ll \alpha/ \log \left(1/\alpha\right)$, there exists a
    ~$\left(2^{\left(1-\alpha\right) n},2^{\left(\alpha-\Theta\left(\delta \log\left(1/\alpha\right)\right)\right)n},\delta,n\right)$-subset code for all large enough~$n$.
\end{corollary}

\subsection{Linear Subset Codes}

\begin{definition}
    A subset code $\mathcal{S}$ is called \emph{linear} if every subset~$S_i \in \mathcal{S}$ is an affine subspace of~$\mathbb{F}_2^n.$
\end{definition}

We next show that contrary to standard error correcting codes, in the case of subset codes linear constructions are not asymptotically optimal.

\begin{lemma}\label{lem:linearsubsetcodes}
    For any~$\alpha,\delta >0$ there is no linear $\left(2^{\left(1-\alpha\right) n},2^{\left(\alpha - \Theta\left(\delta \log \left(\delta / \alpha\right)\right)\right)n},\delta,n\right)$-subset code for all large enough~$n$.
\end{lemma}
\begin{proof}
    Assume~$\mathcal{S}$ was such a linear subset code.
    Consider some~$S_i \in \mathcal{S}$, as it is an affine subspace we have~$S_i=u+V$ where~$u\in\mathbb{F}_2^n$ is some vector and~$V$ is a linear subspace of~$\mathbb{F}_2^n$ of exactly~$\log |S_i|$ dimensions. 
    As such, there exists a subset~$I\subset [n]$ of~$\log |S_i|$ coordinates that are \emph{shattered} by~$V$ (and~$S_i$), that is, the projection of~$S_i$ to these coordinates is surjective. 
    Therefore, the size of~$S_i + B(\delta n / 2)$ is at least~$|S_i| \cdot {n-\log |S_i| \choose \leq \delta n / 2}$; That is because for every possible binary assignment to the coordinates of~$I$, there is a vector in~$S_i$ respecting it --- and thus, at least~${n-\log |S_i| \choose \leq \delta n / 2}$ such vectors in~$S_i + B(\delta n / 2)$.
    Finally, we observe that
    \[
    |S_i|\cdot {n-\log |S_i| \choose \leq \delta n / 2} \geq
    2^{\left(\alpha- \Theta\left(\delta \log \left(\delta / \alpha\right)\right)\right)n} \cdot {\alpha n \choose \leq \delta n / 2} >
    2^{\alpha n}
    ,
    \]
    and thus there are~$i\neq j$ such that~$S_i + B(\delta n / 2)$ intersects~$S_j + B(\delta n / 2)$ --- contradicting~$\mathcal{S}$ being a subset code of distance~$\delta$.
\end{proof}

To illustrate the parameters, consider for example fixing some~$\alpha \in (0,1)$ and taking~$\delta\rightarrow 0$; For general subset codes, we have seen that for~$K=2^{\alpha n}$ the optimal size of each subset is~$T=2^{\left(1-\alpha - \Theta\left(\delta\right)\right)n}$; 
On the other hand, for linear subset codes this optimal size is~$T=2^{\left(1-\alpha -H\left(\delta\right)\right)n}$, which does not improve the total size of~$T\cdot K$ from the bound for the size of standard error correcting codes.

\section{Upper Bounds}\label{sec:upper}

We emulate the proof for linear codes with some more technical complications. 

\begin{theorem}\label{thm:upper}
    For every $(R,\varepsilon)$-unbounded code we have $R\leq 1-\Omega\left(\sqrt{{\varepsilon}}\right)$, if the code is also linear then we have~$R\leq 1-\Omega\left(\sqrt{{\varepsilon \log \left(1/\varepsilon\right)}}\right)$.
\end{theorem}

Let $x=(x_1,x_2, \ldots, x_n, \ldots )$ be a message vector chosen uniformly at random. 
Denote by~$H(i):=H\left(P_i \circ C \left(x\right)\right)$ the entropy of the first~$i$ bits of the code-word of the uniformly chosen message~$x$.
Notice that this definition coincides with that of the linear case if~$C$ is linear. We observe that it still holds that~$H(0)=0$ and~$H(i+1)\leq H(i)+1$ for every~$i$, hence also~$H(i)\leq i$.
For every~$i$ we again denote by~$r(i):=i-H(i)\geq 0$ the \emph{redundancy} of the~$i$-prefix of the code~$C$.
We also follow with an analog to Lemma~\ref{lem:r_upper}.

\begin{lemma}\label{lem:r_upper_nonlin}
    Let~$C$ be a~$(R,\varepsilon)$-unbounded code.
    For every~$i\geq k_0$ we have~$r(i)\leq (1-R)i + 1$.
\end{lemma}
\begin{proof}
    By definition, the first~$i$ bits of a code-word of~$C$ are enough to recover the first~$k:=\lfloor Ri \rfloor$ bits of the message. In particular, the function~$P_k \circ C^{-1}$ is well-defined and we have~$\left(P_k \circ C^{-1}\right) \left(C\left(x\right)[:i]\right) = x[:k]$ for every~$x$.
    We hence have
    \[
    H(i)=H\left(C \left(x\right)[:i]\right) \geq 
    H\left(\left(P_k \circ C^{-1}\right)\left(C \left(x\right)[:i]\right)\right) =
    H\left(x[:k]\right)=k,
    \]
    where the first inequality is known as the \emph{data processing inequality} and the last equality is simply the entropy of a uniform random variable over~$2^k$ elements.
    We conclude that
    \[
    r(i) = i - H(i) \leq i - k \leq i - Ri + 1
    .
    \]
\end{proof}

\begin{lemma}\label{lem:sub}
    Let~$C$ be a~$(R,\varepsilon)$-unbounded code.
    For every~$i,j\in \mathbb{N}$ with~$i\geq k_0$ and~$j\geq\frac{i}{R}$ it holds that
    \[
    r\left(j - \lfloor\frac{j-i}{4}\rfloor\right) - r\left(i-1\right) \geq 
    c \cdot \left(\varepsilon j\right)
    ,
    \]
    for some universal constant~$c>0$.
\end{lemma}
Lemma~\ref{lem:sub} is similar to Lemma~\ref{lem:linearsub}, albeit its proof is substantially more technical. In this proof we make use of the bounds on subset codes from Section~\ref{sec:subset_codes}.

\begin{definition}\label{def:heavy}
    We say that a message prefix~$x_0$ is~$(T,n)$-heavy for a code-word prefix~$c_0$, if
    \[
    \Big\lvert \{ C(x)[:n] \ \mid \ C(x)[:|c_0|] = c_0 \ \text{and} \ x[:|x_0|]=x_0\} \Big\rvert > T
    ,\]
    that is, there are more than~$T$ different code-word prefixes of length~$n$ that begin with~$c_0$ and encode a message that begins with~$x_0$.
    More generally, for a set~$X$ of message prefixes, all of some length~$i$, we say that~$X$ is~$(T,n)$-heavy for a code-word prefix~$c_0$ if
    \[
    \Big\lvert \{ C(x)[:n] \ \mid \ C(x)[:|c_0|] = c_0 \ \text{and} \ x[:i]\in X\} \Big\rvert > T
    .\]
\end{definition}

\begin{lemma}\label{lem:oneheavyset_many}
    Let~$i\geq k_0$, $j\geq i/R$, $\ell < j$, and~$(j-\ell)/8 \leq k \leq (j-\ell)/4$. Let~$c_0$ be a code-word prefix of length~$\ell$.
    Let~$X_1,X_2,\ldots,X_{2^k}$ be~$2^k$ sets of message prefixes of length~$i$ that are pairwise disjoint.
    Then, not all of the sets~$X_p$ are~$\left(T,j\right)$-heavy for~$c_0$, for~$T:=2^{\left(1-\gamma\right)\left(j-\ell\right)-k}$ and some~$\gamma = \Theta\left( \frac{\varepsilon j}{j-\ell} \right)$.
\end{lemma}
\begin{proof}
    Assume the contrary, thus every set~$X_p$ for~$1\leq p \leq 2^k$ is~$(T,j)$-heavy for~$c_0$.
    For each~$p$, by definition of~$(T,j)$-heaviness, we know that there are more than~$T$ distinct code-prefixes~$C(x)[:j]$ of length~$j$ such that~$C(x)[:\ell]=c_0$ and~$x[:i]\in X_p$; Let~$S_p$ be the set of the suffixes~$C(x)[\ell+1 : j]$ of those code-prefixes.
    The sets~$\{S_p\}_{p=1}^{2^k}$ each contain more than~$T$ vectors of length~$m:=j-\ell$.
    By Corollary~\ref{cor:no_subsetcode}, there is no~$\left(2^k, T, \frac{\varepsilon j}{m}, m\right)$-subset code, as we chose~$T=2^{m-k-\Theta\left(\frac{\varepsilon j}{m}\right)m}$. 
    Therefore, there must be~$c_1\in S_{p_1}$ and~$c_2\in S_{p_2}$ such that~$p_1\neq p_2$, and~$|c_1-c_2|\leq \left(\frac{\varepsilon j}{m}\right) m = \varepsilon j$.
    We conclude that there are messages~$x_1',x_2'$ such that~$x_1'[:i]\in X_{p_1}, x_2'[:i]\in X_{p_2}$, hence~$x_1'[:i]\neq x_2'[:i]$, and also $$|C(x_1')[:j]-C(x_2')[:j]| = |C(x_1')[:\ell]-C(x_2')[:\ell]| + |C(x_1')[\ell+1:j]-C(x_2')[\ell+1:j]| = |c_0-c_0|+|c_1-c_2| \leq \varepsilon j,$$
    which is a contradiction to~$C$ being a~$(R,\varepsilon)$-unbounded code.
\end{proof}

\begin{corollary}\label{lem:oneheavy}
    Let~$i\geq k_0$, $j\geq i/R$, $\ell < j$, and~$(j-\ell)/8 \leq k \leq (j-\ell)/4$. Let~$c_0$ be a code-word prefix of length~$\ell$.
    There are less than~$2^k$ message-prefixes~$\{x_p\}$ of length~$i$ that are~$\left(T,j\right)$-heavy for~$c_0$, for~$T$ as defined in Lemma~\ref{lem:oneheavyset_many}.
\end{corollary}
\begin{proof}
    If there are~$2^k$ distinct such code-prefixes~$x_1,\ldots,x_{2^k}$, we apply Lemma~\ref{lem:oneheavyset_many} with~$X_p:=\{x_p\}$ and get a contradiction.
\end{proof}

\begin{corollary}\label{lem:nonheavy}
    Let~$i\geq k_0$, $j\geq i/R$, $\ell < j$, and~$(j-\ell)/8 \leq k \leq (j-\ell)/4$. Let~$c_0$ be a code-word prefix of length~$\ell$.
    Define~$T$ as in Lemma~\ref{lem:oneheavyset_many}.
    Let~$X$ be the set of all message prefixes of length~$i$ which are \emph{not}~$\left(T,j\right)$-heavy for~$c_0$.
    Then,~$X$ is not~$\left(2^{k+1}\cdot T,j\right)$-heavy for~$c_0$.
\end{corollary}
\begin{proof}
    Pick an arbitrary order~$x_1,x_2,\ldots,x_{|X|}$ of the prefixes in~$X$.
    Denote for~$1\leq p \leq |X|$ by~$X_p := \{x_1,x_2,\ldots,x_p\}$ the subset of~$X$ containing the first~$p$ prefixes in the order.
    Denote by~$t_p$ the largest integer~$t$ for which~$X_p$ is~$(t,n)$-heavy for~$c_0$, we let~$t_0:=-1$.
    As each~$x_i$ is not~$(T,j)$-heavy for~$c_0$, it holds that~$t_{p+1}-t_p \leq T-1$ for every~$1\leq p < |X|$.
    Assume by a way of contradiction that~$X=X_{|X|}$ is~$\left(2^{k+1}\cdot T,j\right)$-heavy for~$c_0$; That is, $t_{|X|} \geq 2^{k+1}\cdot T$. 
    
    Hence, we may greedily partition the sequence~$x_1,\ldots,x_{|X|}$ into~$2^k$ consecutive subsequences with each being between~$T$ and~$2T-1$ heavy in the following manner: Let~$q_1,q_2,\ldots,q_{2^k}$ be defined as~$q_1$ being the first index~$r$ for which~$t_r\geq T$, and then iteratively~$q_p$ being the first index~$r$ for which~$t_{r} - t_{q_{p-1}} \geq T$.
    We observe that the process ends as~$t_{p_q}\leq (2T-1) \cdot q$ for every~$q$.
    We also observe that for every~$q$, the set~$X_{p_q} \setminus X_{p_{q-1}}$ is~$(T,j)$-heavy for~$c_0$, and that these~$2^k$ sets are disjoint. 
    We therefore get a contradiction by applying Lemma~\ref{lem:oneheavyset_many}.
\end{proof}

\begin{proof}[Proof of Lemma~\ref{lem:sub}]
    Denote by~$\ell := i - \lfloor\frac{j-i}{4}\rfloor$, by~$m:=j-\ell$, and by~$k:=\lfloor\frac{j-i}{4}\rfloor -1$. 
    We note that~$(j-i)\approx \frac{4}{5} (j-\ell)$ and hence~$\frac{1}{8}(j-\ell)\leq k \leq \frac{1}{4} (j-\ell)$.
    Let~$x$ be a uniformly drawn message, define the random variables~$C_0:=C(x)[:\ell], C_1:=C(x)[\ell+1:j]$; We will estimate the conditional entropy ~$H\left(C_1 \ \mid \ C_0\right)$.
    Let~$T$ be as defined in the statement of Lemma~\ref{lem:oneheavyset_many}.
    Denote by~$B$ the indicator variable of the probabilistic event that~$x[:i]$ is~$(T,j)$-heavy for any code-prefix~$c_0$ of length~$\ell$.
    By Corollary~\ref{lem:oneheavy}, there are at most~$2^k$ prefixes~$x[:i]$ that are~$(T,j)$-heavy for each specific~$c_0$, and thus~$\Pr[B=1]\leq 2^{\ell + k - i}\leq\frac{1}{2}$.
    For any possible assignment~$c_0$ to~$C_0$, and conditioned on the event~$B=0$, Corollary~\ref{lem:nonheavy} tells us that the support of~$C_1$ is of size less than~$2^{k+1}\cdot T$.
    Hence,~$H\left(C_1 \ \mid \ C_0=c_0, \ B=0\right) < \log \left(2^{k+1}\cdot T\right)$.
    As the previous holds for any~$c_0$, we also get~$H\left(C_1 \ \mid \ C_0, \ B=0\right) < \log \left(2^{k+1}\cdot T\right)$.
    As the support of~$C_1$ is always of size at most~$2^m$, we also have~$H\left(C_1 \ \mid \ C_0, \ B=1\right) \leq m$.
    Therefore,
    \begin{align*}
    H\left(C_1 \ \mid \ C_0, B\right) & = 
    \Pr[B=1] \cdot H\left(C_1 \ \mid \ C_0, B=1\right) +
    \Pr[B=0] \cdot H\left(C_1 \ \mid \ C_0, B=0\right) \\&<
    \Pr[B=1] \cdot m + \Pr[B=0] \cdot \log \left(2^{k+1}\cdot T\right) \\& =
    \Pr[B=1] \cdot m + \Pr[B=0] \cdot \left(1 + \left(1-\Theta\left( \frac{\varepsilon j}{m} \right)\right)m \right) \\& =
    \Pr[B=1] \cdot m + \Pr[B=0] \cdot \left(m - \Theta\left(\varepsilon j\right) \right) \\& =
    m - \Pr[B=0] \cdot \Theta\left(\varepsilon j\right).
    \end{align*}
    We further notice that
    \begin{align*}
    H\left(C_1 \ \mid \ C_0\right) \leq 
    H\left(C_1, B \ \mid \ C_0\right) &= 
    H\left(B \ \mid \ C_0\right) + H\left(C_1 \ \mid \ C_0, B\right) \\&\leq
    H\left(B\right) + H\left(C_1 \ \mid \ C_0, B\right) \\&\leq
    1 + H\left(C_1 \ \mid \ C_0, B\right)
    .
    \end{align*}
    Finally, we deduce
    \begin{align*}
        r(j)-r(\ell) &= \left(j - H\left(j\right)\right) - \left(\ell - H\left(\ell\right)\right) \\ &=
        m - \left(H\left(j\right) - H\left(\ell\right)\right) \\ &=
        m - \left(H\left(C_0, C_1\right) - H\left(C_0\right)\right) \\ &=
        m - H\left(C_1 \ \mid \ C_0\right) \\ &\geq
        m - H\left(C_1 \ \mid \ C_0, B\right) - 1 \\ &\geq
        \Pr[B=0] \cdot \Theta\left(\varepsilon j\right) - 1 \\&\geq
        \Omega\left(\varepsilon j\right)
    \end{align*}
\end{proof}

We finally wrap up and the proof of Theorem~\ref{thm:upper}
\begin{proof}[Proof of Theorem~\ref{thm:upper}]
The proof that~$R\leq 1- \Omega(\varepsilon)$ is essentially identical to the proof of Theorem~\ref{thm:upperlin} where Lemma~$\ref{lem:r_upper}$ is replaced with Lemma~\ref{lem:r_upper_nonlin}, and Lemma~\ref{lem:linearsub} is replaced with Lemma~\ref{lem:sub}.

    Let~$n',n\in \mathbb{N}$ be large enough integers, we think of~$n$ as much larger than~$n'$, and of~$n'$ as a large constant. When we write~$o(1)$ or~$\omega(1)$ throughout the proof, it is with respect to~$n'\rightarrow\infty$ and~$\frac{n}{n'}\rightarrow \infty$. We define the sequence~$n_0,\ldots n_k$ recursively by~$n_0:=n$, then $n_{k+1} := \lfloor Rn_k \rfloor - \lfloor\frac{n_k-\lfloor Rn_k \rfloor}{4}\rfloor$ for every~$k\geq 0$, and finally~$K$ is the largest~$k$ such that~$n_k > n'$.
For every~$0\leq k < K$ by using Lemma~\ref{lem:sub} with~$j=n_k,i=\lfloor Rn_k \rfloor$, we have
    \begin{align*}
    r\left(n_k\right) - r\left(n_{k+1}\right) &\geq 
    c \cdot \varepsilon n_k
    .
    \end{align*}

Denote by~$\bar{R} := 1-R$. 
For every~$k$, we have~$n_{k+1} \geq \left(1-\frac{5}{4}\bar{R}\right)n_k - 2$. 
By iterative application of the previous inequality, we also have~$n_k \geq \left(1-\frac{5}{4}\bar{R}\right)^k n_0 - \frac{8}{5\bar{R}}$. This also implies that~$K= \omega(1)$.
Consider the following telescopic summation,
\begin{align*}
r\left(n_0\right) - r\left(n_{K}\right) 
=
\sum_{k=0}^{K-1} \left(r\left(n_k\right) - r\left(n_{k+1}\right)\right)
&\geq
c \sum_{k=0}^{K-1} \varepsilon n_k \\
&=
c \cdot \varepsilon \sum_{k=0}^{K-1} n_k\\
&\geq
\left(1-o\left(1\right)\right) c \cdot \varepsilon \sum_{k=0}^{K-1} \left(1-\frac{5}{4}\bar{R}\right)^k n_0 \\
&=
\left(1-o\left(1\right)\right) c \cdot \varepsilon n \cdot \frac{1}{5\bar{R}/4}
.
\end{align*}
We finally observe that~$r(n_K)\geq 0$, and that by Lemma~\ref{lem:r_upper_nonlin} we also have~$r(n_0)\leq \bar{R} n+1$.
We conclude that~$\bar{R} \geq \frac{4c}{5} \cdot \varepsilon/\bar{R}$, and in particular that~$\bar{R} \geq \Omega\left(\sqrt{\varepsilon}\right)$.

For linear codes, we can use the improved bounds for \emph{linear} subset codes to replace~$\varepsilon$ with~$\varepsilon \log \left(1/\varepsilon\right)$ in the proof.
Consider the sets \[
S_{c_0,x_0}:=\Big\lvert \{ C(x)[:n] \ \mid \ C(x)[:|c_0|] = c_0 \ \text{and} \ x[:|x_0|]=x_0\} \Big\rvert
,
\]
as defined in Definition~\ref{def:heavy}.
When~$C$ is linear, every set~$S_{c_0,x_0}$ is an affine subset of~$\mathbb{F}_2^n$. Furthermore, for a fixed~$c_0$ every set~$S_{c_0,x_0}$ that is not empty is exactly the same size.
Thus, in the proof of Corollary~\ref{lem:oneheavy} we may use Lemma~\ref{lem:linearsubsetcodes} instead of Corollary~\ref{cor:no_subsetcode} and thus set~$\gamma = \Theta\left( \frac{\varepsilon j}{j-\ell} \log \left( \frac{j-\ell}{\varepsilon j} \right) \right) = \Theta\left( \frac{\varepsilon \log\left(1/\varepsilon\right) j}{j-\ell} \right)$ instead of~$\gamma = \Theta\left( \frac{\varepsilon j}{j-\ell} \right)$ in the definition of~$T$.
For the proof of Corollary~\ref{lem:nonheavy} we recall that if there is any non-empty non-heavy~$S_{c_0,x_0}$ for a fixed~$c_0$, then all of these are of the same size. If the number of non-empty such sets is~$\leq 2^k$ then we just think of them all as heavy, otherwise we can use Lemma~\ref{lem:linearsubsetcodes} again and obtain the same~$T$.
Therefore, in the case of a linear~$C$ we end up with~$\bar{R} \geq \Omega\left(\sqrt{\varepsilon \log \left(1/\varepsilon\right)}\right)$.

\end{proof}

\section{Improved Construction}\label{sec:improved_const}
In this section we improve the construction of unbounded codes by using the (non-linear) subset codes of Section~\ref{sec:subset_codes}.
The other building block we use is standard error correcting codes, in particular, we use \emph{systematic codes} or \emph{checksum bits}. A systematic code is an error correcting code in which the codeword contains the message itself. It is well known that any linear error correcting code can be converted into a systematic code of the same rate and distance~\cite{blahut2003algebraic}.
In particular, the following is standard.
\begin{fact}
    For every~$\delta>0$ and large enough~$n$ there exists a linear map~$\text{CS}_\delta:\mathbb{F}_2^n \rightarrow \mathbb{F}_2^{H(\delta)n}$ such that the concatenation~$\left(x||\text{CS}_\delta(x)\right)$ is an ECC with distance~$\delta$.
\end{fact}

Using checksum bits, a natural construction for unbounded codes is the following: Repeatedly transmit a certain number of new message bits, and afterwards transmit a checksum~$CS_\varepsilon$ on the entire prefix of the code beforehand. Right at the end of the checksum bits, the codeword prefix is a true ECC of distance~$\varepsilon$. On the other hand, if we consider a prefix of the codeword that ends right before a checksum, then the entire last chunk of new message bits must be ignored as we are unable to resolve any corruption within it. If the size of the code is~$n$ before such an iteration, and we add~$\alpha n$ new message bits then add the~$\approx H(\varepsilon)n$ checksum bits, then the rate of the code is at most~$\min\left(\frac{\alpha}{\alpha+H\left(\varepsilon\right)}, 1-\alpha\right)$ where the former term comes from the code bits we ``waste" on the checksums and the latter term comes from the last chunk of new message bits we might have to ignore. These two terms equalize at rate~$1-\sqrt{H(\varepsilon)}$.

Our construction of unbounded codes uses checksums and subset-codes in interleave. We remind the reader that a subset code can be viewed as an injective encoding~$\bar{C}:[2^{k_1}]\times[2^{k_2}]\rightarrow \mathbb{F}_2^n$ that has the following two properties: (i) Due to~$e$ being injective, from~$e(x,y)$ we can recover~$x$ and~$y$; (ii) Due to the properties of the code, even if we add errors to~$e(x,y)$ we can still recover~$x$.
In particular, if we use subset-codes to encode the new message bits instead of writing them explicitly, then even before seeing a checksum we would be able to recover \emph{some} (e.g.,~$k_1$) of the new message bits despite possible corruptions. After seeing the checksum, we would be able to recover all of the new message bits (e.g.,~$k_1+k_2$).

\begin{theorem}\label{thm:improv_const}
    For every small enough~$\varepsilon> 0$ and~$R<1-\Omega\left(\sqrt{\varepsilon\log\log \frac{1}{\varepsilon}}\right)$, there exists a $(R,\varepsilon)$-unbounded code.
\end{theorem}

Our construction is iterative. We start by encoding the first $k_0$ (constant) bits of the message with a standard ECC of distance $\varepsilon$ and set $k=k_0$. Then we show how we can take an unbounded code prefix of length~$k$ and extend it by roughly $\sqrt{\varepsilon}\log \nicefrac{1}{\varepsilon} k$ bits. Fix $R=1-\Omega(\sqrt{\varepsilon \log \log {\nicefrac{1}{\varepsilon}}})$.
\begin{lemma}
    Let $C_0:\mathbb{F}_2^k\rightarrow\mathbb{F}_2^{\frac{k}{R}}$ be a code that admits properties of an $(R, \varepsilon)$ unbounded code for $k_0 \leq i\leq k$ and furthermore has relative distance $2\varepsilon$ in its entirety. Let $k_1= k+k\cdot\sqrt{\varepsilon}\log(\nicefrac{1}{\varepsilon})$ then we can extend~$C_0$ into a longer code $C:\mathbb{F}_2^{k_1}\rightarrow \mathbb{F}_2^{\frac{k_1}{R}}$ that admits both of these properties as well.
\end{lemma}
\begin{proof}
    We construct the new code as a concatenation $C_0 || C_1 || CS_{2\varepsilon}$, where $C_0$ is the shorter code we start with, $CS_{2\varepsilon}$ are $O(H(2\varepsilon) k_1)$ parity bits on~$C_0||C_1$ that will ensure that the relative distance of the whole code is at least $2\varepsilon$, and we next detail the construction of~$C_1$.
    The code~$C_1$ encodes $k_1-k=k\sqrt{\varepsilon}\log(\nicefrac{1}{\varepsilon})$ new message bits.
    We partition these message bits into  $\ell=\frac{\log \nicefrac{1}{\varepsilon}}{\sqrt{\log \log\nicefrac{1}{\varepsilon}}}$ consecutive
    blocks~$B_1,B_2,\ldots,B_\ell$ each of size $s=k\sqrt{\varepsilon \log \log \nicefrac{1}{\varepsilon}}$.
    The last block,~$B_\ell$, we further partition into $\ell-1$ sub-blocks~$B_\ell = b_1||b_2||\ldots||b_{\ell-1}$ of size~$\frac{s}{\ell}\approx k \frac{\sqrt{\varepsilon} \log \log \nicefrac{1}{\varepsilon}}{\log \nicefrac{1}{\varepsilon}}$ each.
    Each pair~$(B_j,b_j)$ of a block and a corresponding sub-block will be encoded together using a subset code.
    Let $\bar{C}:\mathbb{F}_2^s\times \mathbb{F}_2^{s/\ell} \rightarrow \mathbb{F}_2^r$ be a $\left(2^{s}, 2^{s/\ell}, 2\sqrt{\frac{\varepsilon}{\log \log \nicefrac{1}{\varepsilon}}} , r\right)$ subset code. By Corollary~\ref{cor:subset} we have such a subset code for $r=s\cdot\left(1+\nicefrac{1}{\ell}+ \Theta\left(\sqrt{\frac{\varepsilon}{\log \log \nicefrac{1}{\varepsilon}}} \cdot \log\left(\ell\right)\right)\right)=s\cdot\left(1+\nicefrac{1}{\ell}+ \Theta\left(\sqrt{\varepsilon \log \log \nicefrac{1}{\varepsilon}}\right)\right)$.  
    We then define~$C_1 := \bar{C}_1 || \bar{C}_2 || \ldots || \bar{C}_{\ell-1}$ where~$\bar{C}_j := \bar{C}(B_j,b_j)$. 
    See Figure~\ref{fig:blocks} for an illustration of how the new message bits are partitioned into blocks and sub-blocks, and which of these are used in every subset code.

    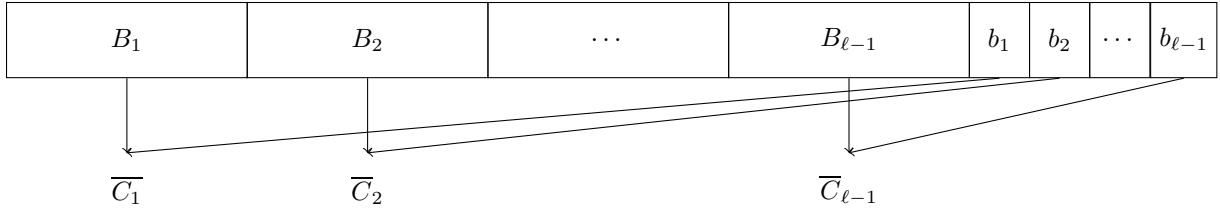
\begin{figure}
\centering
\begin{tikzpicture}
    % Nodes for the top row (large boxes)
    \node[draw, rectangle, minimum width=3.2cm, minimum height=1cm] (box1) at (0, 0) {$B_1$};
    \node[draw, rectangle, minimum width=3.2cm, minimum height=1cm] (box2) at (3.2, 0) {$B_2$};
    \node[draw, rectangle, minimum width=3.2cm, minimum height=1cm] (box_dots) at (6.4, 0) {$\dots$};
    \node[draw, rectangle, minimum width=3.2cm, minimum height=1cm] (box_s) at (9.6, 0) {$B_{\ell-1}$};
    
    % Smaller boxes within the last large box
    \node[draw, rectangle, minimum width=0.8cm, minimum height=1cm] (box_s1) at (11.6, 0) {$b_1$};
    \node[draw, rectangle, minimum width=0.8cm, minimum height=1cm] (box_s2) at (12.4, 0) {$b_2$};
    \node[draw, rectangle, minimum width=0.8cm, minimum height=1cm] (box_dots) at (13.2, 0) {$\dots$};
    \node[draw, rectangle, minimum width=0.8cm, minimum height=1cm] (box_s3) at (14.05, 0) {$b_{\ell-1}$};
    
    % Labels below
    \node at (0, -2) {$\overline{C_1}$};
    \node at (3.2, -2) {$\overline{C}_2$};
    \node at (9.6, -2) {$\overline{C}_{\ell-1}$};

    % Arrows from top boxes to labels
    \draw[->] (box1) -- (0, -1.5);
    \draw[->] (box2) -- (3.2, -1.5);
    \draw[->] (box_s) -- (9.6, -1.5);
    
    % Straight arrows from sub-boxes to labels
    \draw[->] (box_s1.south) -- (0, -1.5);
    \draw[->] (box_s2.south) -- (3.2, -1.5);
    \draw[->] (box_s3.south) -- (9.6, -1.5);
    
    % Curved arrow to label C_s
    %\draw[->] (box2) to[out=-30, in=180] (7.5, -1.5);

\end{tikzpicture}
\label{fig:blocks}
\caption{Partition of the new message bits into blocks and sub-blocks, and their participation in the subset codes.}
\end{figure}

    We first observe that while we encode~$k\sqrt{\varepsilon}\log(\nicefrac{1}{\varepsilon})$ new message bits, we extend the code-word by \begin{align*}
    (\ell-1)\cdot r + O(H(2\varepsilon)) &\leq
    \left(\ell-1\right) \cdot \left(\left(s+\frac{s}{\ell}\right)\cdot\left(1+ \Theta\left(\sqrt{\varepsilon \log \log \nicefrac{1}{\varepsilon}}\right)\right)\right) +  O(H(2\varepsilon)) \\
    &\leq
    k\sqrt{\varepsilon}  \log \nicefrac{1}{\varepsilon} \cdot \left(1 + \Theta\left(\sqrt{\varepsilon \log \log \nicefrac{1}{\varepsilon}}\right)\right)
    \end{align*} 
    bits. Hence, the overall rate is as we desire, as well as the relative distance of the entire code (due to the checksum). It is thus only left to verify that strict prefixes of~$C_0 || C_1 || CS_{2\varepsilon}$ (that are not prefixes of~$C_0$) also satisfy the distance property.
    We observe that the size of every~$\bar{C}_i$ is~$r$ and the total number of corrupted bits is bounded by~$\varepsilon k_1$ and therefore the relative error within each subset code is at most~$\frac{\varepsilon k_1}{r} < 2\sqrt{\frac{\varepsilon}{\log \log \nicefrac{1}{\varepsilon}}}$. Hence, for every~$\bar{C}_j$ that is fully contained in the prefix we are able to recover~$B_j$ despite the errors (but not~$b_j$). This is the reason for the ordering of the new message bits: at the end of each~$\bar{C}_j$ we recover~$B_j$, and only after the final checksum we recover all~$b_j$'s (that is~$B_\ell$) as well.
    Suppose that our prefix ends in the middle of some~$\bar{C}_j$, we will calculate how many message bits that were encoded we are unable to recover: First, as we are in the middle of a subset code we have no guarantees for it, so we can not recover~$B_j$, for every previous~$\bar{C}_i$ for~$i<j$ we are able to recover~$B_i$. Thus, the length of the prefix we have minus the length of~$C_0$ is~$<j\cdot r$, and the number of message bits we recover is~$(j-1)s$. Thus, the length of the prefix of~$C_1$ we see minus the number of new message bits we can recover is at most \[
    r + (j-1)(r-s) \leq
    r + \ell \left(s\cdot\left(\nicefrac{1}{\ell}+ \Theta\left(\sqrt{\varepsilon \log \log \nicefrac{1}{\varepsilon}}\right)\right)\right) = O(s) = 
    O\left(k\sqrt{\varepsilon \log \log \nicefrac{1}{\varepsilon}}\right)
    .
    \]
\end{proof}

We conclude the proof of Theorem~\ref{thm:improv_const} by applying this extension lemma iteratively.

\section{Summary and Open Problems}\label{sec:open}
We introduced and initiated the study of Unbounded Error Correcting Codes. 
Several natural questions remain open.
Primarily, the gap between~$\Omega\left(\sqrt{\varepsilon}\right)$ and~$O\left(\sqrt{\varepsilon\log\log\left(1/\varepsilon\right)}\right)$ in the rate of an optimal code remains open.
Furthermore, many of the questions that were studied in the context of standard ECCs are relevant for unbounded ECCs as well: Can we make the best constructions explicit? Can we have fast encoding and decoding algorithms?
A natural starting point will be an explicit construction for subset codes, as the one we present takes exponential time.

\subsection{Alphabet Size}
In the vast majority of this paper we focus on the binary alphabet~$\Sigma=\Gamma=\mathbb{F}_2$, which is equivalent to the case of an alphabet of any arbitrary constant size~$|\Sigma|,|\Gamma|=O(1)$.
Nonetheless, the same questions we present may also be asked for an alphabet size that is related to~$\varepsilon$.
In standard Error Correcting Codes, the optimal rate of~$1-R=\Theta(\varepsilon \log(1/\varepsilon))$ as~$\varepsilon\rightarrow 0$ is refined to~$1-R=\Theta\left(\varepsilon \left(\log_q\left(1/\varepsilon\right)+1\right)\right)$ when the alphabet size~$q:=|\Sigma|$ is taken into account.
In our construction of the linear code of Section~\ref{sec:const}, the bound is similarly refined to~$1-R=O\left(\sqrt{\varepsilon \left(\log_q\left(1/\varepsilon\right)+1\right)}\right)$ when the dependence on~$q$ is considered. 
In particular, when~$q=\Omega\left(\frac{1}{\varepsilon}\right)$ we obtain a linear code with rate~$1-R=O\left(\sqrt{\varepsilon}\right)$.
On the other hand, the rate upper bound for linear codes in Section~\ref{sec:upperlin} is independent of the alphabet size and thus shows~$1-R\geq \sqrt{\varepsilon}$ for any~$q$, which is tight for~$q=\Omega\left(\frac{1}{\varepsilon}\right)$.

\subsection*{Acknowledgments}
Or Zamir's research is supported in part by the Israel Science Foundation, Grant No. 1593/24, and by the Blavatnik foundation.\\
Klim Efremenko's research is supported by the European Research Council (ERC), Grant No. 949707. 

\bibliography{main}
\bibliographystyle{alpha}

\appendix

\section{Proof of Theorem~\ref{thm:const_rand}}\label{appendix:rand_errors}
\begin{proof}
Our construction is similar to the adversarial case. Each code symbol is a random linear combination of a prefix of the message. That is, we pick random coefficients $a_{i,j}$ and set
\[
C(x_1,x_2\ldots )[j]:=\sum_{i=1}^{r_0 \cdot j} a_{i,j} x_i ,
\]
for some $R<r_0<1-h(3\varepsilon)$ to be chosen later.  
Our decoding procedure is to output the closest codeword in hamming distance, we next show that its corresponding message is consistent with a long prefix of the original message.
\begin{claim}
   Fix~$j$. With positive probability over the choice of all coefficients $a_{i,j}$ the following holds: For every $x\neq y$ such that $k\leq Rj$ is the minimal for which $x[k] \neq y[k]$, we have
 $d_h(C(x)[1,\ldots j], C(y)[1,\ldots j]) \geq 3\varepsilon (j-\frac{k}{r_0})$.
\end{claim}
\begin{proof}
    
Let us first fix $j$ and $k$. Set $n_0= j-\frac{k}{r_0}$. Note that $n_0\geq j-\frac{R}{r_0}j=j(1-R/r_0)$.  The claim holds iff for every $x$ such that $x[1,\ldots k-1]= 0$ and $x[k]=1$ it holds that $wt(C(x)[1\ldots j]) \geq 3\varepsilon n_0$. 
Next note that for any such $x$, $C(x)[\frac{k}{r_0},\ldots, j]$ is a random vector.
Also note that $L= \{C(x)[1,\ldots j]: x[0,\ldots k-1]=0, x[k]=1 \} $ is an affine space of dimension at most $j\cdot r_0- k= n_0\cdot r_0$.  Thus the probability for some vector in $L$ to be of weight at most  $3\varepsilon n_0$ is at most $2^{-n_0}|B(2\varepsilon, n_0)|$ . Thus by union bound any vector in $L$ will have such a weight with probability at most $2^{-n_0+ r_0 n_0}|B(3\varepsilon n_0, n_0)|\leq 2^{-n_0(1-r_0-h(2\varepsilon) -o(1))}$. Thus, if $r_0 < 1-h(3\varepsilon)$, this will be exponentially small. Now note that since  $n_0\geq j(1-R/r_0)$ for a fixed $j$, the sum of this probability over all $k$ will be at most  $\frac{1}{1-r_0-h(2\varepsilon)}2^{-j(1-\frac{R}{r_0})}$. Thus for a large enough $k_0$, the summation over all $j>k_0$ will be less than 1. 
\end{proof}
Now let $\eta=(\eta_1, \eta_2 \ldots \eta_j)$ be the noise vector. From a standard Chrenoff argument, it follows that the probability that for some $ i \leq \frac{R}{r_0}j$ the sub-vector $\eta[i,\ldots j]$ has a relative weight more than $1.5\varepsilon$ is at most exponentially small in $j$. 
Now let us show that for every $\eta=(\eta_1, \eta_2 \ldots \eta_j)$ that does not satisfy this 
the decoding of $C(x)+\eta$ will return $x[1,\ldots Rj]$. 
Let $y$ be different from $x$ in first $Rj$ coordinates. Let $k$ be the first coordinate where they are different.  Then $C(x), C(y)$ are equal in first $\frac{k}{r_0}$ coordinates; thus, noise in these coordinates will not make any difference to prefer $x$ or $y$. Since by the Claim $C(x), C(y)$ differ in $3\varepsilon$ fraction of the remaining coordinates and since relative noise in these coordinates less than half of this $C(x)$ will be closer to the $C(x)+\eta$ that $C(y)$.  
\end{proof}

\end{document}